\documentclass[aps,prl,twocolumn,superscriptaddress,a4paper,nopacs]{revtex4}

\usepackage{graphicx}

\usepackage[colorlinks=true,citecolor=blue,linkcolor=magenta]{hyperref}
\hypersetup{
pdfauthor = {P. Maletinsky},
colorlinks = true, linkcolor = blue, urlcolor=blue, bookmarksnumbered =  true}

\newcommand{\bra}[1]{\left<#1\right|}
\newcommand{\ket}[1]{\left|#1\right>}

\begin{document}

\title{Strong mechanical driving of a single electron spin}

\author{A. Barfuss}
\altaffiliation{These authors contributed equally to this work}
\affiliation{Department of Physics, University of Basel, Klingelbergstrasse 82, CH-4056 Basel, Switzerland}
\author{J. Teissier}
\altaffiliation{These authors contributed equally to this work}
\affiliation{Department of Physics, University of Basel, Klingelbergstrasse 82, CH-4056 Basel, Switzerland}
\author{E. Neu}
\affiliation{Department of Physics, University of Basel, Klingelbergstrasse 82, CH-4056 Basel, Switzerland}
\author{A. Nunnenkamp}
\affiliation{Cavendish Laboratory, University of Cambridge, JJ Thomson Ave., Cambridge CB3 0HE, UK}
\author{P. Maletinsky}
\email{patrick.maletinsky@unibas.ch}
\affiliation{Department of Physics, University of Basel, Klingelbergstrasse 82, CH-4056 Basel, Switzerland}

\date{\today}

\begin{abstract}
Quantum devices for sensing and computing applications require coherent quantum systems which can be manipulated in a fast and robust way\,\cite{Nielsen2000}.
Such quantum control is typically achieved using external electric or magnetic fields which drive the system's orbital\,\cite{Devoret2013,Clarke2008} or spin\,\cite{Hanson2007,Dobrovitski2013} degrees of freedom.
However, most of these approaches require complex and unwieldy 
antenna or gate structures, and with few exceptions\,\cite{Oliver2005,Fuchs2009} are limited to the regime of weak driving.
Here, we present a novel approach to strongly and coherently drive a single electron spin in the solid state using internal strain fields in an integrated quantum device.
Specifically, we study individual Nitrogen-Vacancy (NV) spins embedded in diamond mechanical oscillators and exploit the intrinsic strain coupling between spin and oscillator\,\cite{MacQuarrie2014,Teissier2014, Ovartchaiyapong2014} to strongly drive the spins. 
As hallmarks of the strong driving regime, we directly observe the energy spectrum of the emerging phonon-dressed states\,\cite{Tuorila2010,Silveri2013} and employ our strong, continuous driving for enhancement of the NV spin coherence time\,\cite{Xu2012}.
Our results constitute a first step towards strain-driven, integrated quantum devices and open new perspectives to investigate unexplored regimes of strongly driven multi-level systems\,\cite{Mishra2014,Danon2014} and to study exotic spin dynamics in hybrid spin-oscillator devices\,\cite{Rohr2014,Bennett2013}.
\end{abstract}

\maketitle

The use of crystal strain for the manipulation of single quantum systems (``spins'') in the solid state brings vital advantages compared to established methods relying on electromagnetic fields. Strain fields can be straightforwardly engineered in the solid state and can offer a direct coupling mechanism to embedded quantum systems\,\cite{Hermelin2011,Gustafsson2014,MacQuarrie2014}. Since they are intrinsic to these systems, strain fields are immune to drifts in the coupling strength.
Additionally, strain does not generate spurious stray fields, which are unavoidable with electric or magnetic driving and which can cause unwanted dephasing or heating of the environment. 
Furthermore, coupling quantum systems to strain offers attractive features of fundamental interest. For instance, strain can be used to shuttle information between distant quantum systems\,\cite{Gustafsson2014}, and has been proposed to generate squeezed spin ensembles\,\cite{Bennett2013} or to cool mechanical oscillators to their quantum ground state\,\cite{Wilson-Rae2004}. 
These attractive perspectives for strain-coupled hybrid quantum systems motivated recent studies of the influence of strain on NV centre electronic spins\,\cite{Teissier2014, Ovartchaiyapong2014,MacQuarrie2013} and experiments on strain-induced, coherent driving of large NV spin ensembles\,\cite{MacQuarrie2014}. 
Promoting such experiments to the single spin regime, however, 
remains an outstanding challenge, and would constitute a major first step towards the implementation of integrated, strain-driven quantum systems.


\begin{figure}
\includegraphics[width=8.6cm]{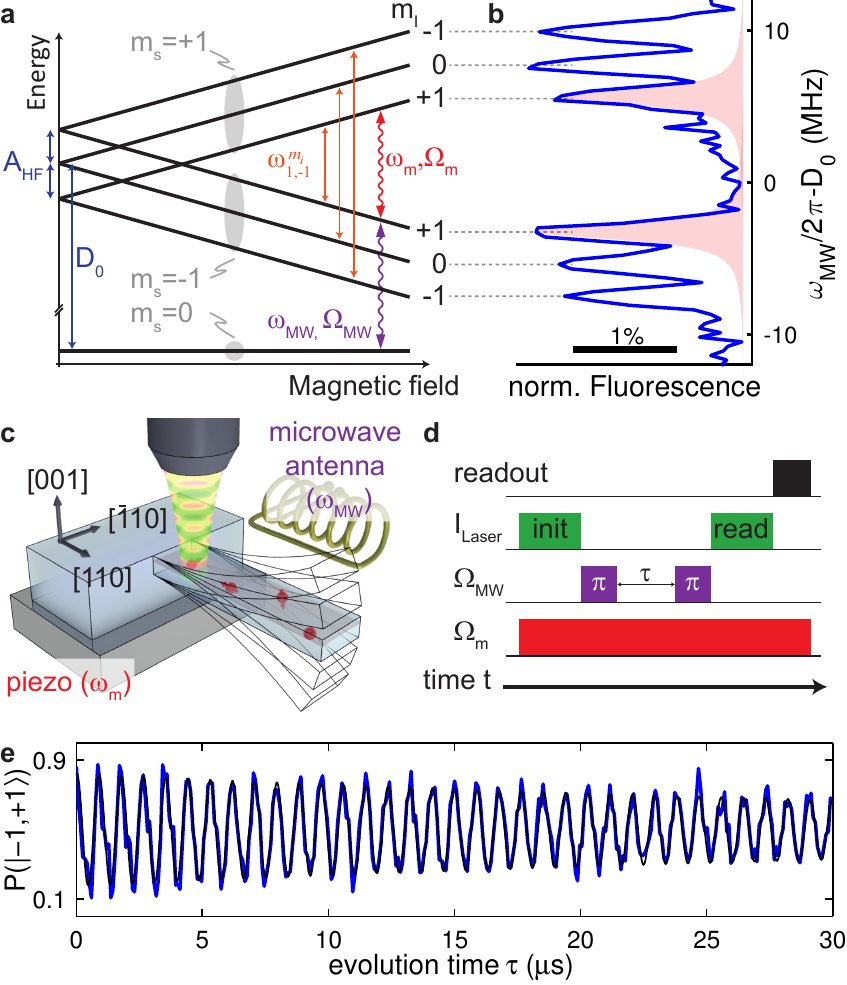} 
\caption{\label{FigStrainDrive} {\bf Experimental setup and strain-induced coherent  spin drive}. {\bf a}~Energy levels of the NV spin as a function of magnetic field applied along the NV axis. Electronic spin states $\ket{m_s=\pm 1}$ each split into three levels due to hyperfine interactions with the NV's $^{14}$N nuclear spin ($I=1$, $A_{\rm HF}=2.18~$MHz). Wavy lines indicate strain (red) and microwave (violet) fields of frequency $\omega$ and strength $\Omega$. {\bf b}~Optically detected electron spin resonance of a single NV centre. {\bf c}~Experimental setup. NV spins (red) are coupled to a cantilever which is resonantly driven at frequency $\omega_m$ by a piezo-element. The NV spin is read out and initialised by green laser light and manipulated by microwave magnetic fields generated by a nearby antenna. {\bf d}~Pulse sequence employed to observe strain-induced Rabi oscillations. {\bf e}~Strain-driven Rabi oscillations. Data (blue) and a fit (black) to damped Rabi oscillations.}
\end{figure}

Here, we demonstrate the coherent manipulation of a single electronic spin using time-periodic, intrinsic strain fields generated in a single-crystalline diamond mechanical oscillator. We show that strain allows us to manipulate the spin in the strong driving regime, where the spin manipulation frequency significantly exceeds the energy splitting between the two involved spin states, and we employ this strong drive to protect the spin from environmental decoherence. 
Our experiments were performed on electronic spins in individual Nitrogen-Vacancy (NV) 
lattice point defect centres, embedded in single-crystalline diamond cantilevers.
The negatively charged NV centres we studied have a spin $S=1$ ground-state with basis states \{$\ket{0}$, $\ket{-1}$, $\ket{+1}$\}, where $\ket{m_s}$ is an eigenstate of the spin operator $\hat{S}_z$ along the NV's symmetry axis, $z$ (Fig.\,\ref{FigStrainDrive}a). The energy difference between $\ket{\pm1}$ and $\ket{0}$ is given by the zero-field splitting $D_0=2.87~$GHz. The levels $\ket{\pm1}$ are split by $2\gamma_{\rm NV}B_{\rm NV}$ (with $\gamma_{\rm NV}=2.8~$MHz/G) in a magnetic field $B_{\rm NV}$ applied along $z$. Hyperfine interactions between the NV's electron and $^{14}$N nuclear spin ($I=1$ and quantum number $m_I$) further split the NV spin levels by an energy $A_{\rm HF}=2.18~$MHz into states $\ket{m_s, m_I}$ (Fig.\,\ref{FigStrainDrive}a)\,\cite{Doherty2012}.
In our experiments, we use optical excitation and fluorescence detection to initialise and optically read out the NV spin\,\cite{Gruber1997} with a homebuilt confocal optical microscope\,\cite{Teissier2014} (Methods). 
Furthermore, we use microwave magnetic fields to perform optically detected electron spin resonance (ESR) (Fig.\,\ref{FigStrainDrive}b) and manipulate the NV's electronic spin states. 

Coherent strain driving of NV spins is based on the sensitive response of the NV spin states to strain in the diamond host lattice. 
For uniaxial strain applied transverse to the NV axis, the corresponding strain-coupling Hamiltonian takes the form\,\cite{Doherty2012, Bennett2013}
\begin{equation}
\label{eqnHcoupl}
H_{\rm str,\perp}=-h g_0^\perp \left(\hat{a} + \hat{a}^\dag\right) \left(S_+^2 + S_-^2\right).
\end{equation}
Here, $h$ is Planck's constant, $g_0^\perp$ the transverse single-phonon strain coupling strength and $\hat{S}_{+(-)}$ and $\hat{a}^\dag$($\hat{a}$) are the raising (lowering) operators for spin and phonons, respectively. 
Transverse strain thus leads to a direct coupling of the two electronic spin states $\ket{-1}$ and $\ket{+1}$\,\cite{Teissier2014} and in the case of near-resonant, time-varying (AC) strain, can coherently drive the transitions $\ket{-1, m_I}\leftrightarrow \ket{+1, m_I}$\,\cite{MacQuarrie2014}. 
For a classical (coherent) phonon field at frequency $\omega_m/2\pi$, Eq.\,(\ref{eqnHcoupl}) can be written as $h \Omega_m/2\pi \cos{(\omega_m t)} \left(S_+^2 + S_-^2\right)$, where $\Omega_m=g_0^\perp x_c/x_{ZPF}$ describes the amplitude of the strain drive, with $x_{ZPF}$ and $x_{c}$ the cantilever's zero-point fluctuation and peak amplitude, respectively.
Interestingly, strain drives a dipole-forbidden transition ($\Delta m_s=2$), which would be inaccessible e.g.~using microwave fields. 

In order to generate and control a sizeable AC strain field for efficient coherent spin driving, we employed a mechanical resonator in form of a singly-clamped, single-crystalline diamond cantilever\,\cite{Teissier2014}, in which the NV centre is directly embedded (Fig.\,\ref{FigStrainDrive}c and Methods). The cantilever was actuated at its mechanical resonance frequency $\omega_m/2\pi=6.83\pm0.02~$MHz using a piezo-element placed nearby the sample. We controlled the detuning between the mechanical oscillator and the $\ket{-1}\leftrightarrow\ket{+1}$ spin transition by applying an adjustable external magnetic field $B_{\rm NV}$ along the NV axis (Fig.\,\ref{FigStrainDrive}a and Methods). 


\begin{figure}
\includegraphics[width=8.6cm]{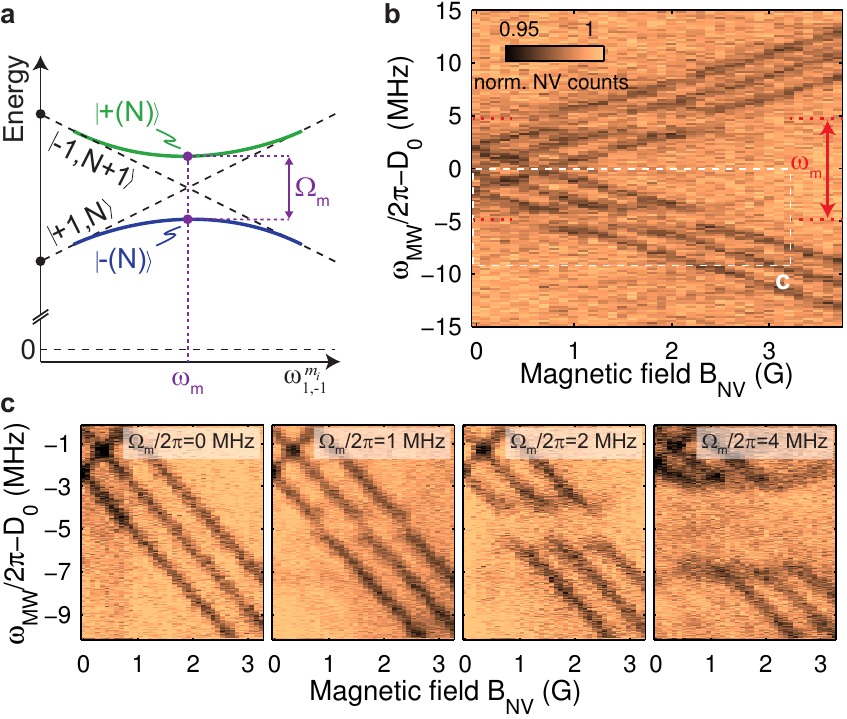}
\caption{\label{FigACStark} {\bf Mechanically induced Autler-Townes effect probed by microwave spectroscopy.} {\bf a} Eigenenergies of the joint spin-phonon system with basis-states $\ket{m_s; N}$ as a function of the spin splitting $\omega_{1,-1}^{m_I}$. Strain couples $\ket{1; N}$ and $\ket{-1; N+1}$ and, whenever the resonance condition $\omega_{1,-1}^{m_I}=\omega_m$ is fulfilled, leads to new eigenstates $\ket{+(N)}$ and $\ket{-(N)}$ with energy splitting  $\Omega_m$ (see text). {\bf b} Microwave spectroscopy of phonon-dressed states using a weak microwave probe at $\omega_{\rm MW}$. For $m_I=-1,0$ and $1$, resonance is separately established at $B_{\rm NV}\sim0.9,1.6$ and $2.3~$G. {\bf c} Dependance of the energy gap between $\ket{+(N)}$ and $\ket{-(N)}$ on mechanical driving strength. As expected, the gap scales linearly with $\Omega_m$ for each hyperfine state. Data was recorded over a parameter range indicated by white dashed lines in {\bf b}.}
\end{figure}

To demonstrate coherent NV spin manipulation using resonant AC strain fields, we first performed strain-driven Rabi oscillations between $\ket{-1}$ and $\ket{+1}$ for a given hyperfine manifold (here, $m_I=1)$. 
To that end, we initialised the NV in $\ket{-1,1}$ by applying an appropriate sequence of laser and microwave pulses (Fig.\,\ref{FigStrainDrive}d, inset). 
We then let the NV spin evolve for a variable time $\tau$, under the influence of the coherent AC strain field generated by constantly exciting the cantilever at 
a fixed peak amplitude 
$x_{c}\sim 100~$nm (Supplementary Information). 
After this evolution, we measured the resulting population in $\ket{-1,1}$ with a pulse sequence analogous to our initialisation protocol. 
As expected, we observe strain-induced Rabi oscillations (Fig.\,\ref{FigStrainDrive}d) for which we find a Rabi frequency $\Omega_m/2\pi=1.14\pm0.01~$MHz and 
hardly any damping over the $30~\mu$s observation time.
Importantly and in contrast to a recent study on NV ensembles\,\cite{MacQuarrie2014}, this damping timescale is not limited by ensemble-averaging, since our experiment was performed on a single NV spin.

We obtain further insight into the strength and dynamics of our coherent strain-driving mechanism from ESR spectroscopy of the strain-coupled NV spin states, $\ket{+1}$ and $\ket{-1}$. For this, we employed a weak microwave tone at frequency $\omega_{\rm MW}$ to probe the $\ket{0}\leftrightarrow\ket{\pm1}$ transitions as a function of $B_{\rm NV}$ in the presence of the coherent strain field (Fig.\,\ref{FigACStark}b). This field has a striking effect on the NV's ESR spectrum in that it induces excitation gaps at frequencies $\omega_{\rm MW}-2\pi D_0=\pm\omega_m/2$, i.e.~at $B_{\rm NV}\sim0.9,1.6$ and $2.3~$G. 
At these values of $B_{\rm NV}$ the AC strain field in the cantilever is resonant with a given hyperfine transition, i.e.~the energy splitting $\hbar\omega_{1,-1}^{m_I}$ between $\ket{-1, m_I}$ and $\ket{+1, m_I}$ equals $\hbar\omega_m$. 
The energy gaps which we observe in the ESR spectra under resonant strain driving are evidence of the Autler-Townes effect\,\cite{Autler1955}, which we here observe for the first time on a single spin in the microwave domain.

The observed Autler-Townes splitting can be understood by considering the joint energetics of the NV spin states and the quantised strain field used to drive the spin\,\cite{Cohen-Tannoudji1992} (Fig.\,\ref{FigACStark}a). The joint basis states $\ket{i; N}$ consist of NV spin states $\ket{i}$ dressed by $N$ phonons in the cantilever. Strain couples $\ket{+1; N}$ to $\ket{-1; N+1}$ 
and leads to new eigenstates $\ket{\pm(N)}$, which anti-cross on resonance, where $\ket{\pm(N)}=(\ket{+1; N}\pm\ket{-1; N+1})/\sqrt{2}$ are split by an energy $\hbar\Omega_m$. 
As expected, this splitting increases linearly with the driving field amplitude (Fig.\,\ref{FigACStark}c), which we control through the strength of piezo excitation.


\begin{figure}
\includegraphics[width=8.6cm]{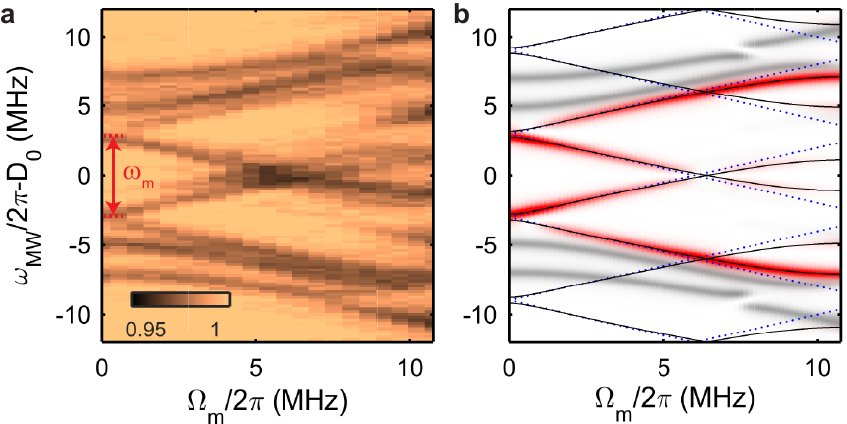}
\caption{\label{FigStrongDrive} {\bf Dressed-state spectroscopy of the strongly driven NV spin.} {\bf a} Microwave spectroscopy of the mechanically driven NV spin at $B_{\rm NV}=1.8~$G (i.e.~$\omega_m=\omega_{1,-1}^{m_I=+1}$) as a function of drive strength $\Omega_{m}$. The resonantly coupled states ($\ket{\pm1, +1}$) at $\omega_{\rm MW}/2\pi-D_0=\pm2.98~$MHz first split linearly with $B_{\rm NV}$ and then evolve in a sequence of crossings and anticrossings with higher-order dressed states, indicative of the strong driving regime. The color scale is identical to the one used in Fig.\,\ref{FigACStark}b. {\bf b} Calculated dressed-state energy spectrum (black lines). Colour scale: transition rates from $\ket{m_s=0}$ to the dressed states obtained by Fermi's Golden rule (Methods). The red-shaded transitions correspond to the hyperfine manifold $m_I=+1$.}
\end{figure}

To investigate the limits of our coherent strain-induced spin driving and study the resulting, strongly driven spin dynamics, we performed detailed dressed-state spectroscopy as a function of drive strength (Fig.\,\ref{FigStrongDrive}a).
To that end, we first  
set $B_{\rm NV}$ such that $\omega_{-1, 1}^{m_I=1}=\omega_m$ 
and then performed microwave ESR spectroscopy for different values of $\Omega_m$. 
For weak driving, $\Omega_m \ll \omega_m$, the dressed states emerging from the resonantly coupled states $\ket{-1, 1}$ and $\ket{+1, 1}$ split linearly with $\Omega_m$.
The linear relationship breaks down for $\Omega_m\gtrsim\omega_{-1, 1}^{m_I=1}$ due to multi-phonon couplings involving states which belong to different sub-spaces spanned by $\ket{\pm(N)}$ and $\ket{\pm(M)}$, with $N\neq M$\,\cite{Cohen-Tannoudji1992}.
This observation is closely linked to the breakdown of the rotating-wave approximation\,\cite{Fuchs2009} and indicates the onset of the strong driving regime we achieve in our experiment.

For even larger Rabi frequencies $\Omega_m$, the dressed states evolve into a characteristic sequence of crossings and anti-crossings. 
The (anti-)crossings occur whenever $\Omega_m=q\omega_m$, with $q$ an odd (even) integer, and are related to symmetries of Hamiltonian\,(\ref{eqnHcoupl})\,\cite{Cohen-Tannoudji1992} (Supplementary Information). 
Our experiment clearly allows us to identify the $q=1$ and $q=2$ (anti-)crossings (Fig.\,\ref{FigStrongDrive}a) and thereby demonstrates that we reside well within the strong driving regime ($\Omega_m>\omega_{-1, 1}^{m_I}$) of a harmonically driven two-level system. We have carried out an extensive numerical analysis (Fig.\,\ref{FigStrongDrive}b and Methods), which shows quantitative agreement with our experimental findings. Our calculation further shows that over our range of experimental parameters, $\Omega_m$ is linear in $x_c$ and reaches a maximum of $\Omega_m^{\rm max}/2\pi\sim10.75~$MHz.

Strong coherent driving can be employed to protect a quantum system from its noisy environment and thereby increase its coherence times\,\cite{Cai2012,Xu2012,Golter2014,Timoney2011}. 
For NV centre spins, decoherence is predominantly caused by environmental magnetic field noise\,\cite{DeLange2010}, which normally couples linearly to the NV spin through the Zeeman Hamiltonian $H_Z=\gamma_{\rm NV}S_zB_{NV}$ (Fig.\,\ref{FigStrainDrive}a). 
Conversely, for the dressed states $\ket{\pm(N)}$ we create by strong coherent driving, $\bra{\pm(N)}H_Z\ket{\pm(N)}=0$ and the lowest order coupling to magnetic fields is only quadratic (Fig.\,\ref{FigACStark}a). These states are thus  less sensitive to magnetic field fluctuations and should exhibit increased coherence times, compared to the undriven NV.


\begin{figure}
\includegraphics[width=8.6cm]{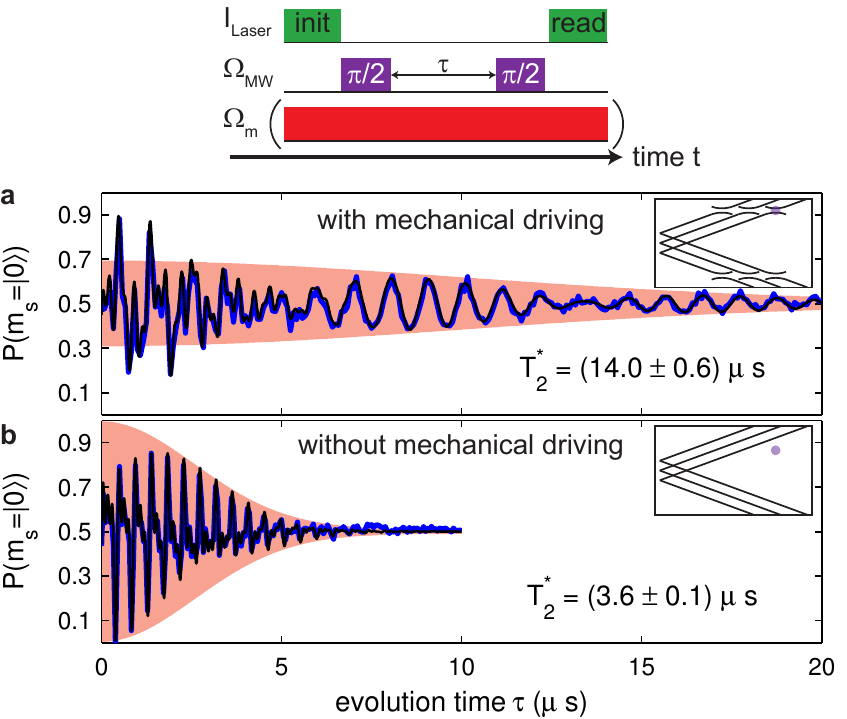}
\caption{\label{FigLinewidth} {\bf Protecting NV spin coherence by strong mechanical driving.} {\bf a} Spin coherence decay of $\ket{\pm(N)}$ (for $m_I=+1$) as measured by Ramsey interferometry between the state $\ket{m_s=0}$ and the $\ket{m_s=+1}-$manifold, using the pulse sequence depicted on the top. The inset illustrates the NV spin's eigenenergies as a function of $B_{NV}$ and indicates the magnetic field and microwave frequencies employed (red dot) with respect to the dressed-state spectrum shown in Fig.\,\ref{FigACStark}b. {\bf b} Measurement of NV spin coherence time in the un-driven case, as determined by Ramsey spectroscopy between $\ket{0,0}$ and $\ket{1,0}$, in the absence of mechanical driving. Inset as in a. The orange envelope indicates the coherence decay extracted from our fit.}
\end{figure}

To demonstrate such coherence enhancement by continuous driving\,\cite{Xu2012}, we performed Ramsey spectroscopy on our strongly strain-driven NV spin and compared the resulting dephasing times $T_2^*$ to the undriven case (Fig.\,\ref{FigLinewidth}). For this, we adjusted $B_{\rm NV}$ such that $\omega_{-1, 1}^{m_I=1}=\omega_m$ and mechanically drove the NV with $\Omega_m/2\pi=1.86~$MHz 
to induce phonon-dressing of the NV. We then used pulsed microwaves to perform Ramsey spectroscopy on the two Autler-Townes split dressed states emerging from $\ket{m_s=+1,m_I=+1}$. 
The resulting coherence signal (Fig.\,\ref{FigLinewidth}a) decays on a timescale of $T_2^*=14.0\pm 0.6~\mu$s and shows beating of two long-lived oscillations at $0.92~$MHz and $1.04~$MHz 
stemming from the two dressed states we address (Supplementary Information).
Compared to the bare NV dephasing time of $T_2^*=3.6\pm0.1~\mu$s (Fig.\,\ref{FigLinewidth}b), this demonstrates a significant enhancement of $T_2^*$ caused by our strong, mechanical drive.

Several factors contribute to the remaining dressed-state decoherence we observe: While our protocol decouples the NV from magnetic field noise, it renders it vulnerable to fluctuations in electric field and strain. Strain field noise is dominated by the cantilever's thermal motion, which can be safely neglected (Supplementary Information). However, our shallow NVs might experience excess dephasing from surface electric fields\,\cite{Dolde2011}, whose precise noise spectral density and influence on NV spin decoherence is subject to current research. 
An additional residual dephasing might be induced by the remaining second-order coupling of magnetic field fluctuations to the NV spin\,\cite{Ithier2005} and could be suppressed further by increasing $\Omega_m$.
A detailed analysis of these dephasing processes, along with their dependance on $\Omega_m$ will be subject to further study.


Our approach to strong coherent strain-driving of a single electronic spin will have implications far beyond the coherence protection and dressed-state spectroscopy that we have demonstrated in this work.
By combining our strain-drive with coherent microwave spin manipulation, our NV spin forms an inverted three-level ``$\Delta$''-system, on which all three possible spin transitions can be coherently addressed. This setting is known to lead to unconventional spin dynamics\,\cite{Yamamoto1998, Buckle1986}, which here could be observed on a single-spin basis and exploited for sensing and quantum manipulation of our hybrid device.
Strain-induced AC Stark shifts can furthermore be employed to dynamically tune\,\cite{Jundt2008} the energies of the NV hyperfine states -- an attractive perspective for the use of $^{14}$N nuclear spins as quantum memories\,\cite{Fuchs2011}.
The decoherence protection by strong driving we demonstrated will be impactful for any quantum technology where pulsed decoupling protocols cannot be employed (such as DC electric field sensing). 
Further studies of the remaining decoherence processes under strong driving, which remain largely unexplored until now, offer another exciting avenue to be pursued in the future. 
On a more far-reaching perspective, our experiments lay the foundation for exploiting diamond-based hybrid spin-oscillator systems for quantum information processing and sensing, where our system forms an ideal platform for implementing proposed schemes for spin-induced phonon cooling and lasing\,\cite{Kepesidis2013} or oscillator-induced spin squeezing\,\cite{Bennett2013}. 

\section{Methods}

{\bf Sample fabrication:} Our cantilevers consist of single-crystalline, ultra-pure [001]-oriented diamond (Element Six, ``electronic grade''), are aligned with the [110] crystal direction and have dimensions in the range of $(0.2-1)\times3.5\times(15-25)~\mu$m$^3$ for thickness, width and length, respectively.
The fabrication process is based on recently established top-down diamond nanofabrication techniques\,\cite{Maletinsky2012}. In particular, we use electron beam lithography at $30~$keV to pattern etch masks for our cantilevers into a negative tone electron beam resist (FOX-16 from Dow Corning, spun to a thickness of $\sim500~$nm onto the sample). The developed pattern directly acts as an etch mask and is transferred into the diamond surface by using an inductively coupled plasma reactive ion etcher (ICP-RIE, Sentech SI 500). To create cantilevers with vertical sidewalls, we use a plasma containing $50\%$ argon and $50\%$ oxygen (gas flux $50~$sccm each). The plasma is run at $1.3~$Pa pressure, $500~$W ICP source power, and $200~$W bias power.
NV centres in our cantilevers were created prior to nanofabrication by $^{14}$N ion implantation with dose, energy and sample tilt of $10^{10}~$cm$^{-2}$, $12~$keV and $0^\circ$, respectively. Based on numerical simulations (using the ``SRIM'' software package), this yields an estimated implantation depth of $\sim17~$nm. To create NV centres, we annealed our samples at high vacuum ($\lesssim10^{-6}~$mbar) in a sequence of temperature steps at $400~^\circ$C ($4~$hours), $800~^\circ$C (2 hours) and $1200~^\circ$C (2 hours)\,\cite{Chu2014}. 

{\bf Experimental setup:} Experiments are performed in a homebuilt confocal microscope setup at room temperature and at atmospheric pressure. A $532~$nm laser (NovaPro $532$-$300$) is coupled into the confocal system through a dichroic mirror (Semrock LM01-552-25). A microscope objective (Olympus XLMFLN40x) is used to focus the laser light onto the sample, which is placed on a micropositioner (Attocube ANSxyz100).  
Red fluorescence photons are collected by the same microscope objective, transmitted through the dichroic mirror and coupled  into a single mode optical fibre (Thorlabs SM600), which acts as a pinhole for confocal detection. Photons are detected using an avalanche photodiode (Laser Components Count-250C) in Geiger mode. Scan control and data acquisition (photon counting) are achieved using a digital acquisition card (NI-6733). The microwave signal for spin manipulation is generated by a SRS SG384 signal generator, amplified by a Minicircuit ZHL-42W+ amplifier and delivered to the sample using a homebuilt near-field microwave antenna. Laser, microwave and detection signals were gated using microwave switches (Minicircuit Switch ZASWA-2-50DR+), which were controlled through digital pulses generated by a fast pulse generator (SpinCore PulseBlasterESR-PRO). Gating of the laser is achieved using a double pass acoustic optical modulator (Crystal Technologies 3200-146).
Mechanical excitation of the cantilevers was performed with a piezoelectric element placed directly below the sample. The excitation signal for the piezo was generated with a signal generator (Agilent 3320A). A three-axis magnetic field was generated by three homebuilt coil pairs driven by constant-current sources (Agilent E3644A). 

{\bf Measurement procedure and error bars:} ESR measurements were performed using a pulsed ESR scheme\,\cite{Dreau2011}, where the NV spin is first initialised in $\ket{m_s=0}$ using green laser excitation, then driven by a short microwave ``$\pi$-pulse'' of length $\tau$ (i.e., a pulse such that $\Omega_{\rm MW}\tau=\pi$) and finally read out using a second green laser pulse. Compared to conventional, continuous-wave ESR, this scheme has the advantage of avoiding power broadening of the ESR lines by green laser light and was therefore employed throughout this work.

Our experiments were performed on three different NV centres in three different cantilevers: Data in Figs.\,\ref{FigStrainDrive} and\,\ref{FigLinewidth} were obtained on NV $\#1$, while Figs.\,\ref{FigACStark} and\,\ref{FigStrongDrive} were recorded on NVs~$\#2$ and $\#3$, respectively. NVs~$\#1-3$ all showed slightly different values of $D_0$ due to variations in static local strain and transverse magnetic fields. The zero-field splittings for these three NVs were $D_0=2.870~$GHz, $2.871~$GHz and $2.8725~$GHz, respectively. The values of $\omega_m$ for the cantilevers of NVs~$\#1-3$ were $\omega_m/2\pi=6.83, 5.99$ and $5.95~$MHz, respectively.

Throughout this paper, errors represent $95~\%$ confidence intervals for the nonlinear least-squares parameter estimates to our experimental data. The only exception is the mechanical resonance frequency $\omega_m$, where error bars represent the linewidth of the cantilever resonance curves, which we measured optically in separate experiments. The actual error bars in determining $\omega_m$ are significantly smaller than the linewidth and do not influence the findings presented in this paper.

{\bf Simulations:} Following Ref.~\cite{Silveri2013} we employ Floquet theory to treat the time dependence of the strain-induced spin driving, $\hat{H}_m = \hbar\Omega_m \cos \omega_m t \left(S_+^2 + S_-^2\right)$, beyond rotating-wave approximation (RWA), since it is expected to break down in the strong driving limit $\Omega_m>\omega_{-1, 1}^{m_I=1}$. The key idea here is to map the Hamiltonian with periodic time dependence on an infinite-dimensional, but time-independent Floquet Hamiltonian $\mathcal{H}_F$. We can then solve the eigenvalue problem $\mathcal{H}_F \ket{u_j} = \hbar \omega_j \ket{u_j}$ with standard methods to obtain quasi-energies $\hbar \omega_j$ and corresponding eigenvectors $\ket{u_j}$.

Treating the weak microwave drive up to second order in drive strength we find the rate for the system to leave the initial state with Fermi's golden rule as \cite{Silveri2013}
\begin{equation}
\mathcal{P} = \frac{1}{\hbar^2} \sum_{i,f} \frac{\gamma_{fi} |\bra{u_f}\hat{H}_{\rm MW}\ket{u_i}|^2}{(\omega_f -\omega_i - \omega_P)^2+\frac{\gamma_{fi}^2}{4}},
\end{equation}
with the microwave driving Hamiltonian $\hat{H}_{\rm MW} = \sum_{m_I} \hbar \Omega_{\rm MW} (\ket{+1,m_I}\bra{0,m_I} + \ket{-1,m_I}\bra{0,m_I} + \rm{H.c.})$ with drive frequency $\Omega_{\rm MW}$.
For the simulations shown in Fig.~\ref{FigStrongDrive}b we assumed an initial state $\ket{u_i} = \ket{m_s=0,m_I}$ and linewidths $\gamma_{fi} = \gamma=1~$MHz, and summed the result incoherently over all nuclear spin quantum numbers $m_I\in\{-1,0,1\}$.

\section{Acknowledgements}
We thank V. Jacques for fruitful discussions, P. Appel for initial assistance with nanofabrication and L. Thiel for support with the experiment control software. 
We gratefully acknowledge financial support from SNI; NCCR QSIT; SNF grants 200021\_143697; and EU FP7 grant 611143 (DIADEMS). 
AN is supported by the Royal Society.

%


\bibliographystyle{apsrev4-1}
\bibliography{BibCoherentStrainDriveALL}

\newpage

\section*{Supplementary Information}

\section*{S1. Crossings and anticrossings and linearity of $x_c$ with $\Omega_m$ in Fig.3 of the main text}

{\em Crossings and anticrossings:} For an explanation of the series of crossings and anti-crossings observed in Fig.\,3 of the main text, it is convenient to introduce the new quantum number $\eta(m_s,N)=-1^{m_s/2+N-1/2}$, in analogy to the corresponding treatment of dressed states of atoms driven by radio frequency fields\,\cite{Cohen-Tannoudji1992}. 
Since $H_{\rm str,\perp}$ changes $N$ by $\pm1$ and $m_s$ by $\pm2$, it changes $m_s/2+N$ by $0$ or $2$ and therefore conserves $\eta(m_s,N)$ as defined above. It is then straightforward to verify that the crossings and anti-crossings correspond to degenerate and near-degenerate dressed states with unequal and equal values of $\eta(m_s,N)$.

{\em Linearity of $x_c$ with $\Omega_m$:} The experimental data presented in Fig.\,3a of the main text was obtained by taking microwave ESR spectra for increasing values of the driving strength of the cantilever, which we set by the voltage $V_p$ used to drive the excitation piezo. The data shows near-perfect agreement with the calculated dressed-state spectra presented in Fig.\,3b. From this match, we conclude that $\Omega_m \propto V_p$ and since $V_p \propto x_c$\,\cite{Teissier2014}, the linearity stated above follows. This observation has relevance for the strain-induced coherent manipulation of NV spins in that potential nonlinearities could be an obstacle to deterministic, coherent strain-driving.

\section*{S2. Limits to decoupling by mechanical driving}

{\em Thermal noise of the cantilever:} The total root-mean-square (RMS) thermal noise amplitude of a cantilever amounts to $x_{th}=\sqrt{\frac{k_B T}{k}}$, with $k_B=1.3\cdot10^{-23}~$J/K Boltzmann's constant, $T$ the temperature and $k$ the oscillator's spring constant. For a singly clamped cantilever $k=\frac{Ewt^3}{4l^3}$\,\cite{Sarid1994}, with Youngs modulus $E$, cantilever width $w$, thickness $t$, and length $l$. For our diamond cantilevers with typical dimensions $w\sim2~\mu$m, $t\sim1~\mu$m, $l\sim20~\mu$m and $E=1220~$GPa, we find $k=76~$N/m and therefore $x_{th}\sim10~$pm. 

{\em Temperature fluctuations:} Temperature fluctuations of the diamond sample over the course of the experiment could also lead to damping of the measured Ramsey fringes. This is caused by the non-negligible dependance of the zero field splitting parameter $D_0$ on temperature, which was recently measured to be d$D_0/$dT$=-78~$kHz$/$C$^\circ$\,\cite{Fang2013}. The dephasing of our measured Ramsey fringes (Fig.3 of the main text) occurs at a rate of $\sim 71~$kHz and could thus be explained by temperature variations of  $\sim 1.1~$C$^\circ$.

\section*{S3. Amplitude of cantilever excitation}

We determined the physical amplitude of our cantilever excitation, by observing the reduction in fluorescence of a single NV close to the end of the cantilever as a function of excitation voltage. We have described this method in detail elsewhere\,\cite{Teissier2014} and used it to calibrate the cantilever's mechanical susceptibility $\chi_m$ to the piezo drive amplitude, for which we found $\chi_m=12~$nm/V. For the experiment presented in Fig.1d of the main text, we employed a piezo voltage of $8~$V$_{\rm pp}$, which resulted in the cantilever amplitude $x_c\sim100~$nm, which we quoted in the text.

\section*{S4. Experimental details regarding Ramsey spectroscopy}

For the Ramsey spectroscopy presented in Fig.4a of the main text, we employed the pulse sequence shown on the top of Fig.4 with a pulsed microwave field strength $\Omega_{\rm MW}/2\pi=1.29~$MHz and frequency $\omega_{\rm MW}/2\pi-D_0=6.83/2~$MHz. This value of $\omega_{\rm MW}/2\pi$ corresponds to a spectral position roughly in the center of the two dressed state transitions, i.e. to a symmetrical detuning of $\sim\pm1~$MHz of the microwave to these transitions (see inset in Fig.4a). The microwave Rabi frequency $\Omega_{\rm MW}$ was thereby strong enough to drive both detuned transitions but also induced some unwanted population in $\ket{+1,-1}$ and $\ket{+1,0}$. These populations led to the fast oscillating and highly damped signal visible for $\tau\lesssim5~\mu$s in Fig.4a.

In order to fit the Ramsey data in Fig.4, we used a sum of cosines with Gaussian decays\,\cite{Maze2012} for both dressed states and the uncoupled hyperfine states.
In the un-driven case (Fig.4b) we use the fitting function  
$\exp\left[-(\tau/T_2^*)^2\right]\sum_{m_i}\beta_{m_i}\cos(2\pi\delta_{m_i}\tau+\phi_{m_i})$, 
where $\beta_{m_i}$, $\delta_{m_i}$ and  $\phi_{m_i}$ are the population, microwave-detuning and initial phase for the hyperfine state $\ket{+1,m_I}$.
In the driven case, the dressed state coherences decay on a different timescale as compared to the bare hyperfine states. We therefore employed the fitting function 
$\exp\left[-(\tau/T_{2, DS}^*)^2\right]\beta_{\rm DS}\sum_{i\in\{+,-\}}\cos(2\pi\delta_{DS,i}\tau+\phi_{DS,i})+ \sum_{m_j\in\{-1,0\}}\exp\left[-(\tau/T_{2,m_j}^*)^2\right]\beta_j\cos(2\pi\delta_{m_j}\tau+\phi_{m_j})$, 
where parameters are defined as before and the subscript ``DS'' indicates fit paramters related to dressed states.
Note that, due to the onset of off-resonant dressing of $\ket{+1,-1}$ and $\ket{+1,0}$, these states showed slightly increased decay times as compared to the completely undriven case and in particular, each needed to be fit with its own decay time $T_{2,m_j}^*$. 

\end{document}